\documentstyle[twocolumn,aps,graphicx]{revtex}

\begin{document}
\twocolumn[\hsize\textwidth\columnwidth\hsize\csname
@twocolumnfalse\endcsname

\def\simge{\hspace*{0.2em}\raisebox{0.5ex}{$>$}
     \hspace{-0.8em}\raisebox{-0.3em}{$\sim$}\hspace*{0.2em}}
\def\simle{\hspace*{0.2em}\raisebox{0.5ex}{$<$}
     \hspace{-0.8em}\raisebox{-0.3em}{$\sim$}\hspace*{0.2em}}
\def\bra#1{{\langle#1\vert}}
\def\ket#1{{\vert#1\rangle}}
\def\coeff#1#2{{\scriptstyle{#1\over #2}}}
\def\undertext#1{{$\underline{\hbox{#1}}$}}
\def\hcal#1{{\hbox{\cal #1}}}
\def\sst#1{{\scriptscriptstyle #1}}
\def\eexp#1{{\hbox{e}^{#1}}}
\def\rbra#1{{\langle #1 \vert\!\vert}}
\def\rket#1{{\vert\!\vert #1\rangle}}
\def\lsim{{ <\atop\sim}}
\def\gsim{{ >\atop\sim}}
\def\nubar{{\bar\nu}}
\def\psibar{{\bar\psi}}
\def\Gmu{{G_\mu}}
\def\alr{{A_\sst{LR}}}
\def\wpv{{W^\sst{PV}}}
\def\evec{{\vec e}}
\def\notq{{\not\! q}}
\def\notk{{\not\! k}}
\def\notp{{\not\! p}}
\def\notpp{{\not\! p'}}
\def\notder{{\not\! \partial}}
\def\notcder{{\not\!\! D}}
\def\notA{{\not\!\! A}}
\def\notv{{\not\!\! v}}
\def\Jem{{J_\mu^{em}}}
\def\Jana{{J_{\mu 5}^{anapole}}}
\def\nue{{\nu_e}}
\def\mn{{m_\sst{N}}}
\def\mns{{m^2_\sst{N}}}
\def\me{{m_e}}
\def\mes{{m^2_e}}
\def\mq{{m_q}}
\def\mqs{{m_q^2}}
\def\mz{{M_Z}}
\def\ubar{{\bar u}}
\def\dbar{{\bar d}}
\def\sbar{{\bar s}}
\def\qbar{{\bar q}}
\def\gv{{g_\sst{V}}}
\def\ga{{g_\sst{A}}}
\def\pv{{\vec p}}
\def\pvs{{{\vec p}^{\>2}}}
\def\ppv{{{\vec p}^{\>\prime}}}
\def\ppvs{{{\vec p}^{\>\prime\>2}}}
\def\qv{{\vec q}}
\def\qvs{{{\vec q}^{\>2}}}
\def\xv{{\vec x}}
\def\xpv{{{\vec x}^{\>\prime}}}
\def\yv{{\vec y}}
\def\tauv{{\vec\tau}}
\def\sigv{{\vec\sigma}}
\def\sst#1{{\scriptscriptstyle #1}}
\def\gpnn{{g_{\sst{NN}\pi}}}
\def\grnn{{g_{\sst{NN}\rho}}}
\def\gnnm{{g_\sst{NNM}}}
\def\hnnm{{h_\sst{NNM}}}

\def\xivz{{\xi_\sst{V}^{(0)}}}
\def\xivt{{\xi_\sst{V}^{(3)}}}
\def\xive{{\xi_\sst{V}^{(8)}}}
\def\xiaz{{\xi_\sst{A}^{(0)}}}
\def\xiat{{\xi_\sst{A}^{(3)}}}
\def\xiae{{\xi_\sst{A}^{(8)}}}
\def\xivtez{{\xi_\sst{V}^{T=0}}}
\def\xivteo{{\xi_\sst{V}^{T=1}}}
\def\xiatez{{\xi_\sst{A}^{T=0}}}
\def\xiateo{{\xi_\sst{A}^{T=1}}}
\def\xiva{{\xi_\sst{V,A}}}

\def\rvz{{R_\sst{V}^{(0)}}}
\def\rvt{{R_\sst{V}^{(3)}}}
\def\rve{{R_\sst{V}^{(8)}}}
\def\raz{{R_\sst{A}^{(0)}}}
\def\rat{{R_\sst{A}^{(3)}}}
\def\rae{{R_\sst{A}^{(8)}}}
\def\rvtez{{R_\sst{V}^{T=0}}}
\def\rvteo{{R_\sst{V}^{T=1}}}
\def\ratez{{R_\sst{A}^{T=0}}}
\def\rateo{{R_\sst{A}^{T=1}}}

\def\mro{{m_\rho}}
\def\mks{{m_\sst{K}^2}}
\def\mpi{{m_\pi}}
\def\mpis{{m_\pi^2}}
\def\mom{{m_\omega}}
\def\mphi{{m_\phi}}
\def\Qhat{{\hat Q}}

\def\FOS{{F_1^{(s)}}}
\def\FTS{{F_2^{(s)}}}
\def\GAS{{G_\sst{A}^{(s)}}}
\def\GES{{G_\sst{E}^{(s)}}}
\def\GMS{{G_\sst{M}^{(s)}}}
\def\GATEZ{{G_\sst{A}^{\sst{T}=0}}}
\def\GATEO{{G_\sst{A}^{\sst{T}=1}}}
\def\mdax{{M_\sst{A}}}
\def\mustr{{\mu_s}}
\def\rsstr{{r^2_s}}
\def\rhostr{{\rho_s}}
\def\GEG{{G_\sst{E}^\gamma}}
\def\GEZ{{G_\sst{E}^Z}}
\def\GMG{{G_\sst{M}^\gamma}}
\def\GMZ{{G_\sst{M}^Z}}
\def\GEn{{G_\sst{E}^n}}
\def\GEp{{G_\sst{E}^p}}
\def\GMn{{G_\sst{M}^n}}
\def\GMp{{G_\sst{M}^p}}
\def\GAp{{G_\sst{A}^p}}
\def\GAn{{G_\sst{A}^n}}
\def\GA{{G_\sst{A}}}
\def\GETEZ{{G_\sst{E}^{\sst{T}=0}}}
\def\GETEO{{G_\sst{E}^{\sst{T}=1}}}
\def\GMTEZ{{G_\sst{M}^{\sst{T}=0}}}
\def\GMTEO{{G_\sst{M}^{\sst{T}=1}}}
\def\lamd{{\lambda_\sst{D}^\sst{V}}}
\def\lamn{{\lambda_n}}
\def\lams{{\lambda_\sst{E}^{(s)}}}
\def\bvz{{\beta_\sst{V}^0}}
\def\bvo{{\beta_\sst{V}^1}}
\def\Gdip{{G_\sst{D}^\sst{V}}}
\def\GdipA{{G_\sst{D}^\sst{A}}}
\def\fks{{F_\sst{K}^{(s)}}}
\def\FIS{{F_i^{(s)}}}
\def\fpi{{F_\pi}}
\def\fk{{F_\sst{K}}}

\def\RAp{{R_\sst{A}^p}}
\def\RAn{{R_\sst{A}^n}}
\def\RVp{{R_\sst{V}^p}}
\def\RVn{{R_\sst{V}^n}}
\def\rva{{R_\sst{V,A}}}
\def\xbb{{x_B}}

\def\PR#1{{{\em   Phys. Rev.} {\bf #1} }}
\def\PRC#1{{{\em   Phys. Rev.} {\bf C#1} }}
\def\PRD#1{{{\em   Phys. Rev.} {\bf D#1} }}
\def\PRL#1{{{\em   Phys. Rev. Lett.} {\bf #1} }}
\def\NPA#1{{{\em   Nucl. Phys.} {\bf A#1} }}
\def\NPB#1{{{\em   Nucl. Phys.} {\bf B#1} }}
\def\AoP#1{{{\em   Ann. of Phys.} {\bf #1} }}
\def\PRp#1{{{\em   Phys. Reports} {\bf #1} }}
\def\PLB#1{{{\em   Phys. Lett.} {\bf B#1} }}
\def\ZPA#1{{{\em   Z. f\"ur Phys.} {\bf A#1} }}
\def\ZPC#1{{{\em   Z. f\"ur Phys.} {\bf C#1} }}
\def\etal{{{\em   et al.}}}

\def\delalr{{{delta\alr\over\alr}}}
\def\pbar{{\bar{p}}}
\def\lamchi{{\Lambda_\chi}}


\title{Parity-Violating Electron Scattering as a Probe of Supersymmetry}

\author{A. Kurylov$^{a,b}$, M.J. Ramsey-Musolf$^{a,b}$, and S. Su$^a$
\\[0.3cm]
}
\address{
$^a$ California Institute of Technology,
Pasadena, CA 91125\ USA\\
$^b$ Department of Physics, University of Connecticut, Storrs, CT 06269\ USA
}


\maketitle

\begin{abstract}

We compute the one-loop supersymmetric (SUSY) contributions to the weak
charges of
the electron
($Q_W^e$) and proton ($Q_W^p$) using the Minimal Supersymmetric Standard
Model (MSSM).
These $q^2=0$ vector couplings of the $Z^0$-boson to fermions will be
determined
in two fixed-target,
parity-violating electron scattering experiments. The SUSY loop
contributions to $Q_W^p$ and
$Q_W^e$ can be substantial, leading to several percent corrections to the
Standard Model values
for these quantities. We show that the relative signs of the SUSY loop effects
on $Q_W^e$ and $Q_W^p$
are correlated and positive over nearly all of the MSSM parameter space,
whereas inclusion of
R-parity nonconserving interactions can lead to opposite sign relative shifts
in the weak charges.
Thus, a comparison of $Q_W^p$ and $Q_W^e$ measurements could help
distinguish between different
SUSY scenarios.

\end{abstract}



\vspace{0.3cm}
]

\pagenumbering{arabic}

The search for physics beyond the Standard Model (SM) of electroweak and
strong interactions is
a primary objective for particle and nuclear physics. Historically,
parity-violating (PV)
interactions have played an important role in elucidating the structure of
the electroweak
interaction. In the 1970's,  PV deep inelastic scattering (DIS) measurements
performed at the Stanford Linear Accelerator Center (SLAC) confirmed the SM
prediction for the structure of weak neutral current
interactions\cite{Pre78}. These results were
consistent with a value for the weak mixing angle given by
$\sin^2\theta_W\approx 1/4$, implying
a tiny $V$(electron)$\times A$(quark) neutral current interaction.
Subsequent PV measurements --
performed at both very low scales using atoms as well as at the $Z^0$-pole
in $e^+e^-$
annihilation -- have  been remarkably consistent with the results of the
SLAC DIS
measurement\cite{Pre78}.

More recently, the results of cesium atomic
parity-violation (APV) \cite{Ben99} and deep inelastic $\nu$- ($\bar\nu$-)
nucleus
scattering\cite{Zel02}  have been interpreted as determinations of the
scale-dependence of
$\sin^2\theta_W$. The SM predicts how this parameter will evolve from its
precisely measured
value at the
$Z^0$-pole. The cesium APV and neutrino DIS measurements imply $-2\sigma$
and $+3\sigma$
deviations, respectively, from the predicted evolution of $\sin^2\theta_W$
(defined in
the $\overline{\rm MS}$ scheme). If
conventional atomic or hadron structure effects are ultimately unable to
account for these
discrepancies,  the results of these precision measurements would point
to new physics.

In light of this situation, two new measurements involving polarized
electron scattering
have taken on added interest: PV M\"oller ($ee$) scattering at
SLAC\cite{E158} and elastic,  PV
$ep$ scattering at the Jefferson Lab (JLab)\cite{Qweak}. In the absence of
new physics, both
measurements could be used to determine $\sin^2\theta_W$ at the same scale
($|Q^2|\approx
0.03$ $({\rm GeV}/c)^2$) -- falling between the scales relevant
to the APV and neutrino DIS measurements -- with comparable precision in
each case. Any
significant deviation from the SM prediction for $\sin^2\theta_W$ at this
scale  would provide
striking evidence for new physics, particularly if
both measurements report a deviation. On the other hand, agreement would
imply that the
most likely explanations for the cesium APV and neutrino DIS results are
atomic and
hadron structure effects within the SM.

In this Letter, we analyze the prospective implications of the
parity-violating electron
scattering (PVES) measurements for supersymmetry (SUSY). Although no
supersymmetric
particle has yet been discovered, there exists strong motivation for
believing that SUSY is a
component of the \lq\lq new" Standard Model. For example, the existence of
low-energy SUSY is a prediction of many string theories; it offers a
solution to the
hierarchy problem, providing a mechanism for maintaining the stability of
the electroweak scale
against large radiative corrections; and it results in coupling unification
close to the Planck scale.
In light of such arguments, it is clearly of interest to determine
what insight about SUSY the new PVES measurements might provide.

In the simplest version of SUSY -- the Minimal Supersymmetric Standard
Model (MSSM)\cite{Haber}
-- low-energy precision observables experience SUSY only via tiny loop
effects involving
virtual supersymmetric particles. In the
MSSM, the requirement of baryon minus lepton number ($B-L$)
conservation leads to conservation of the R-parity quantum number,
$P_R=(-1)^{2S+3(B-L)}$, where $S$ denotes spin. Every SM particle has
$P_R=+1$ while
the corresponding superpartner, whose spin differs by $1/2$ unit, has $P_R=-1$.
Conservation of $P_R$ implies that every vertex has an even number of
superpartners.
Consequently, for processes like $ee\to ee$, all superpartners must live in
loops. Such
loops generate corrections -- relative to the SM amplitude -- of order
$(\alpha/\pi)(M/{\tilde M})^2$, where $M$ denotes a SM particle mass and
${\tilde M}$ is a
superpartner mass.  Thus, low-energy experiments must probe an observable
with a precision of
few tenths of a percent  or better in order to discern SUSY loop effects.
Low-energy
charged current experiments have already reached such levels of precision,
and the
corresponding implications of these experiments for the MSSM have been
discussed
elsewhere\cite{kurylov02}.

The leading-order SM contribution to the PV $ee$ and $ep$ asymmetries is
governed by the
tree-level vector coupling of the $Z^0$-boson to these fermions -- the
so-called
\lq\lq weak charge". At tree-level in the SM this coupling is suppressed:
$Q_W^p=-Q_W^e=1-4\sin^2\theta_W\approx 0.1$. One-loop SM electroweak radiative
corrections further reduce this tiny number, leading to the predictions
$Q_W^e=-0.0449$\cite{Mar96,Erl02} and
$Q_W^p=0.0721$\cite{Erl02}. The fortuitous suppression of these couplings
in the SM renders them
more transparent to the possible effects of new physics. Consequently,
experimental precision of order a few percent, rather than a
few tenths of a percent, is needed to probe SUSY loop corrections. As we
show below, the
possible magnitude of these effects may be as large as the proposed
experimental
error bars for the $Q_W^{e}$ and $Q_W^p$ measurements (8\% and 4\%,
respectively
\cite{E158,Qweak}). Moreover,
the relative sign of the effect in both
cases is correlated -- and positive -- over nearly all available SUSY
parameter space. To our
knowledge, this correlation is specific to the MSSM, making it a potential
low-energy signature
of this new physics scenario.

The content of the MSSM has been described in detail elsewhere\cite{Haber},
so we review only a
few features here. The particle spectrum consists of the SM particles and
the corresponding
superpartners: spin-0 sfermions ($\tilde f$), spin-$1/2$
gluinos ($\tilde g$), and
spin-$1/2$ mixtures of Higgsinos and electroweak gauginos, the neutralinos
($\tilde
\chi^0_{1-4}$) and charginos ($\tilde \chi^{\pm}_{1,2}$). In addition, the
Higgs sector of
the MSSM contains two doublets (\lq\lq up"- and \lq\lq down"-types,
respectively), whose
vacuum expectations $v_u$ and $v_d$ are parameterized in terms of
$v=\sqrt{v_u^2+v_d^2}$
and $\tan\beta=v_u/v_d$. Together with the SU(2)$_L$ and U(1)$_Y$ couplings
$g$ and $g'$
respectively, $v$ is determined from $\alpha$, $M_Z$, and the Fermi
constant extracted
from the muon lifetime, $G_\mu$, while $\tan\beta$ remains a free
parameter. The MSSM
also introduces a coupling between the two Higgs doublets characterized by the
dimensionful parameter $\mu$.

Degeneracy between SM particles and their superpartners is lifted by the
SUSY-breaking
Lagrangian, which depends in general on 105 additional parameters. These
include the SUSY-breaking Higgs mass parameters; the
gaugino masses $M_{1,2,3}$; the left- (right-)handed sfermion mass
parameters
$M_{\tilde f_L}^2$ ($M_{\tilde f_R}^2$); and terms which mix ${\tilde f}_L$ and
${\tilde f}_R$ into mass eigenstates ${\tilde f}_{1,2}$.
One expects the magnitude of the SUSY-breaking
parameters to lie
somewhere between the weak scale and $\sim 1$ TeV. Significantly larger
values can
reintroduce the hierarchy problem.

Theoretical models for SUSY-breaking mediation provide
relations among this large set of parameters, generally resulting in only a few
independent parameters at the SUSY-breaking or GUT scale\cite{Kane}.
According to the
model-independent  analysis of Ref. \cite{kurylov02}, however, the
superpartner spectrum implied
by these models conflicts with the combined constraints of low-energy
charged current data,
$M_W$, and the muon anomalous magnetic moment unless one allows for
nonconservation of $P_R$.
Here, we adopt a similar model-independent approach and do not impose any
specific relations
among SUSY-breaking parameters. To our knowledge, no other
model-independent analysis of MSSM
corrections to neutral current observables has appeared in the literature,
nor have the complete
set of corrections to low-energy PV observables been computed previously
(see, e.g.
\cite{erler}).

MSSM loop corrections to elementary $e$-$f$ amplitudes consist of several
topologies.
Contributions to gauge-boson self energies  can be expressed entirely in
terms of the oblique parameters $S$, $T$, and $U$ in the limit that
${\tilde M} \gg M_Z$.
Since present collider limits allow for fairly light superpartners,
however, we do not
work in this limit. Consequently, the corrections arising from the photon
self-energy
($\Pi_{\gamma\gamma}$) and $\gamma$-$Z^0$ mixing tensor ($\Pi_{Z\gamma}$)
contain a residual
$q^2$-dependence not embodied by the oblique parameters.
Vertex corrections, external leg corrections,
and box graphs are process-dependent and cannot be parameterized in any
general way. Note that one must consider corrections to both neutral
current and charged
current amplitudes, since the former are normalized to
$G_\mu$ and since $\sin^2\theta_W$ is
calculated in the MSSM
using $\alpha$, $M_Z$, $G_\mu$ and radiative corrections to these
quantities. We
evaluate these corrections using the modified dimensional reduction
renormalization scheme
(${\overline{\rm DR}}$) \cite{siegel}.

Including these corrections, the general structure for an elementary
$A(e)\times V(f)$
amplitude is
\begin{equation}
{\cal M} = -{G_\mu\over 2\sqrt{2}}Q_W^f\ {\bar e}\gamma_\mu\gamma_5 e\ {\bar
f}\gamma^\mu f
\end{equation}
with
\begin{equation}
Q_W^f = \rho_{PV}\left[2 T_3^f -4\kappa_{PV}
Q_f\sin^2\theta_W\right]+\lambda_f\ \ \ .
\end{equation}
Here, $\rho_{PV}=1+\delta\rho_{PV}$ and $\kappa_{PV}=1+\delta\kappa_{PV}$,
with $\delta\rho_{PV}$
and $\delta\kappa_{PV}$ denoting
a universal set of corrections to PV amplitudes and $\lambda_f$ containing
the effects of
process-specific vertex, external leg, and box graph effects. It is useful
to express
the SUSY contributions to  $\delta\rho_{PV}$ and $\delta\kappa_{PV}$ in
terms of oblique
parameters as well as a residual $q^2$-dependence of $\Pi_{Z\gamma}$
and $\Pi_{\gamma\gamma}$:
\begin{eqnarray}
\delta\rho^{\rm susy}_{PV} & = & {\hat\alpha} T-{\hat\delta}_{VB}^\mu\\
\label{eq:kappapv}
\delta\kappa_{PV}^{\rm susy} & = & -{\hat\alpha}\left({{\hat c}^2\over
{\hat c}^2-{\hat s}^2}
\right) T +{\hat\alpha}\left[{1\over 4{\hat s}^2({\hat c}^2-{\hat
s}^2)}\right] S\\
\nonumber
&&+{{\hat c}\over {\hat s}}\Bigl[  {{\hat\Pi}_{Z\gamma}(q^2)\over q^2}-
{{\hat\Pi}_{Z\gamma}(M_Z^2)\over M_Z^2}\Bigr]\\
\nonumber
&&+\Bigl({{\hat c}^2\over {\hat c}^2-{\hat
s}^2}
\Bigr)\Bigl[{\Delta{\hat\alpha}\over{\hat\alpha}}-
{{\hat\Pi}_{\gamma\gamma}(M_Z^2)\over
M_Z^2}
+{\hat\delta}_{VB}^\mu\Bigr]\ \ \ .
\end{eqnarray}
Here, the hat denotes quantities renormalized in the ${\overline{\rm DR}}$
scheme; ${\hat\delta}_{VB}^\mu$ are the SUSY
vertex, external leg, and box graph corrections to the $\mu$-decay amplitude;
${\hat s}$ (${\hat c}$) is
the sine (cosine) of the weak mixing angle in the ${\overline{\rm DR}}$
scheme defined
at the scale $\mu=M_Z$; and
$\Delta{\hat\alpha}$ is the SUSY contribution to the difference between the
fine structure
constant and the electromagnetic coupling renormalized $\mu=M_Z$. Note that
superpartner loops give ${\hat\Pi}_{Z\gamma}(q^2)\propto q^2$.

In order to evaluate the potential size of SUSY corrections, we
generated  set of $\sim 3000$ different combinations of SUSY-breaking
parameters, chosen
randomly from flat distribution in mass parameters and $\ln\tan\beta$. The
former were
bounded below by present collider limits and bounded above by 1000 GeV,
corresponding
to the ${\cal O}$(TeV) naturalness limit. We also restricted $\tan\beta$ to
lie in the
range $1.4 < \tan\beta < 60$ as required by perturbativity of third
generation quark Yukawa
couplings, and we allowed for left-right mixing among sfermions.  In order
to avoid unacceptably
large flavor-changing neutral currents, we have also assumed no
intergenerational sfermion
mixing.

For each combination of parameters, we evaluate superpartner masses and
mixing angles, which we
then use as inputs for computing the radiative corrections. We also
separately evaluate the
corresponding contributions to the oblique parameters. The latter are
tightly constrained from
precision electroweak data. We rule out any parameter combination leading
to values of $S$ and
$T$ lying outside the present 95\% confidence limit contour for these
quantities. We note that
this procedure is not entirely self-consistent, since we have not evaluated
non-universal MSSM
corrections to other precision electroweak observables before extracting
oblique parameter
constraints. As noted in Ref.
\cite{erler}, where MSSM corrections to $Z$-pole observables were evaluated
using
different models for SUSY-breaking mediation, non-universal effects can be
as large as
oblique corrections. Nevertheless, we expect our procedure to yield a
reasonable estimate
of the oblique parameter constraints.  Since  $S$ and $T$ do not dominate
the low-energy
SUSY corrections (see below), our results depend only gently on the precise
allowed ranges for
these parameters.

For the case of charged current observables, gluino loops can generate
substantial
corrections when the masses and mixing angles for $u$- and $d$-type squarks
are not
identical\cite{kurylov02}. In contrast, gluinos decouple entirely from the
one-loop MSSM
corrections to semi-leptonic neutral current PV observables. Moreover, MSSM
Higgs contributions to
vertex, external leg, and box graph corrections are
negligible due to the small, first- and second-generation Yukawa couplings.
The light Higgs
contribution to gauge boson propagators has already been included via the
oblique parameters,
while the effects of other MSSM Higgs bosons are sufficiently small to be
neglected\cite{haber93}.

In Fig. 1, we plot the shift in the weak charge of the proton, $\delta Q_W^p =
2\delta Q_W^u+ \delta Q_W^d$, versus the corresponding shift in the
electron's weak
charge, $\delta Q_W^e$, normalized to the respective SM values.
The corrections can be as large as $\sim 4\%$ ($Q_W^p$) and $\sim 8\%$
($Q_W^e$) -- roughly the size of the proposed experimental errors.
Generally speaking, the magnitudes of $\delta Q_W^{e,p}$ decrease as SUSY
mass parameters are
increased. The largest effects occur when at least one superpartner is
relatively light.
An exception occurs in the presence of significant mass splitting between
sfermions, which may lead to sizable contributions. However, such weak
isospin-breaking
effects also increase the magnitude of $T$, so their impact is bounded by
oblique
parameter constraints. This consideration has been implemented in
arriving at
Fig. 1. We also observe that the presence or absence of sfermion left-right
mixing
affects the distribution of points, but does not significantly change the
range of
possible corrections. For the situation of no left-right mixing, the points
are more
strongly clustered near the origin. Thus, while corrections of the order of
several percent
are possible in either case, large effects are more likely in the presence of
left-right mixing.

The shifts $\delta Q_W^{e,p}$ are dominated by $\delta\kappa_{PV}^{\rm
susy}$. Non-universal
corrections involving vertex corrections and wavefunction renormalization
experience significant
cancellations, while box graphs are numerically suppressed. We find that
$\delta\kappa_{PV}^{\rm
susy}$ is nearly always negative, corresponding to a reduction in the
effective $\sin^2\theta_W$
for the PVES experiments. Within $\delta\kappa_{PV}^{\rm susy}$
itself, contributions from the various terms in Eq. (\ref{eq:kappapv}) have
comparable
importance, with some degree of cancellation occurring between the effects
of $S$ and $T$. Thus,
the oblique parameter approximation gives a rather poor description of the
MSSM effects on the
weak charges.

As evident from Fig. 1, the relative sign of the corrections to both
$Q_W^p$ and
$Q_W^e$ is nearly always the same and nearly always positive. Since $Q_W^p>0$
($ Q_W^e<0$) in the SM, SUSY loop corrections give $\delta Q_W^p>0$ ($\delta
Q_W^e<0$).
This correlation is significant, since the effects of other new
physics scenarios can display different signatures. For example, for the
general class of $E_6$ neutral gauge bosons (with mass $\lesssim$ 1000
GeV), the effects on
$Q_W^p$ and $Q_W^e$ also correlate, but $\delta Q_W^{e,p}/Q_W^{e,p}$ can
have either sign in this
case\cite{MRM99,Erl02}. Leptoquark interactions, in contrast, would not
lead to discernible
effects in $Q_W^e$ but could induce sizable shifts in $Q_W^p$ \cite{MRM99}.

As a corollary, we also note that the relative importance of SUSY loop
corrections to
the weak charge of heavy nuclei probed with APV is suppressed. The shift in the
nuclear weak charge is given by $\delta Q_W(Z,N) = (2Z+N)\delta Q_W^u +
(2N+Z)\delta
Q_W^d$. Since the  sign of $\delta Q_W^f/Q_W^f$ due to superpartner loops
is nearly
always the same, and since $Q_W^u>0$ and $Q_W^d<0$ in the SM, a strong
cancellation between $\delta Q_W^u$ and $\delta Q_W^d$ occurs in heavy
nuclei. This
cancellation implies that the magnitude of $\delta Q_W(Z,N)/Q_W(Z,N)$ is
generally less than
about 0.2\% for cesium and is equally likely to have either sign. Since the
presently quoted
uncertainty for the cesium nuclear weak charge is
about 0.6\%, the present $\sim 2\sigma$ deviation from the SM prediction
does not
substantially
constrain the SUSY parameter space.

While agreement between experimental values for $Q_W^{e,p}$ and the SM
predictions would
not produce significant new constraints on the MSSM, a deviation of $\sim
2\sigma$ or more
could help distinguish between the MSSM and other new physics scenarios.
For example, a
$+2.5\sigma$ deviation for $Q_W^{e,p}$, along with the cesium APV result,
would be
consistent with an
$E_6$ $Z'$ boson but fairly difficult to accomodate in the MSSM. To
illustrate, we plot
in Fig. 2 the impact of such a result on the MSSM parameter space.
Here, we have assumed no L-R mixing, set $\tan\beta=10$, and taken
$\mu=M_1=M_2/2 =
100$ GeV. The plot indicates the allowed left-handed slepton and squark
masses (${\tilde M}_L$
and ${\tilde M}_Q$), assuming for simplicity a common mass for all
generations. Present collider
searches rule out the dark shaded region, while charged current data
exclude the light shaded
area. The hatched region would be ruled out (at 95\% C.L.) by a large,
positive $\delta Q_W^{p,e}$,
yielding an upper bound ${\tilde M}_Q {\sim 300}$ GeV. Should future
collider
limits exceed this value, then a $Z'$ would be the favored explanation of
of large cesium APV
and $Q_W^{p,e}$ deviations.

Alternately, one may relax the assumption of $P_R$ conservation. Doing
so leads to new interactions of the type ${\tilde e}{\bar e}\nu$ and
${\tilde d}^{\ast}{\bar e}^c u$  with coupling strengths $\lambda_{ijk}$ and
$\lambda^\prime_{ijk}$ (the subscripts refer to generation number). These
interactions give rise
to new tree-level exchange of  sfermions between SM fermions. As discussed
in Ref. \cite{MRM00},
charged current data, $M_W$, and the results of cesium APV can be
accommodated under this
scenario if
$\lambda_{12k}$ and  $\lambda^\prime_{11k}$ are nonzero.  The interior of
the truncated
ellipse in Fig. 1 shows
the possible corrections to $Q_W^p$ and $Q_W^e$, given the constraints from
other electroweak
data.  We observe that the prospective effects of
$P_R$ non-conservation are quite distinct from SUSY loops. The shifts
$\delta Q_W^p/Q_Q^p$ and $\delta Q_W^e/Q_W^e$ have opposite signs over
most of the allowed region, in contrast to the situation for
SUSY loop effects or
$E_6$ $Z'$ bosons. Thus, a comparison of results for the two PVES
experiments could help
determine whether this extension of the MSSM is to be favored over other new
physics scenarios.

We thank R. Carlini, B. Filippone, S. Heinemeyer, R.D. McKeown, and M. Wise
for useful
comments. This work was supported in part by the U.S. Department of Energy and
National Science Foundation.




\begin{figure}
\hspace{0.1in}
\includegraphics[height=2.25in,width=2.5in]{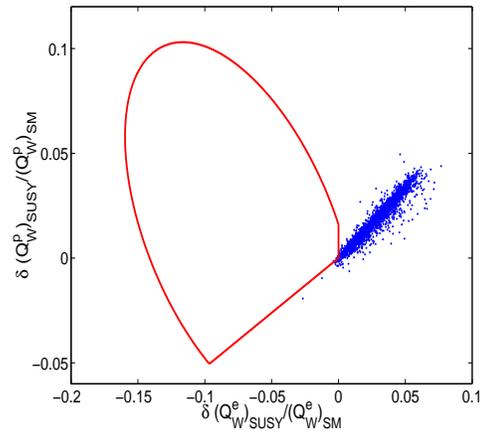}
\caption{\label{Fig1} Relative shifts in electron and proton weak charges
due to
SUSY effects. Dots indicate MSSM loop corrections for $\sim 3000$
randomly-generated
SUSY-breaking parameters. Interior of truncated elliptical region gives
possible shifts
due to $P_R$
nonconserving SUSY interactions (95\% confidence).}
\end{figure}

\begin{figure}
\hspace{0.05in}
\includegraphics[height=2.25in]{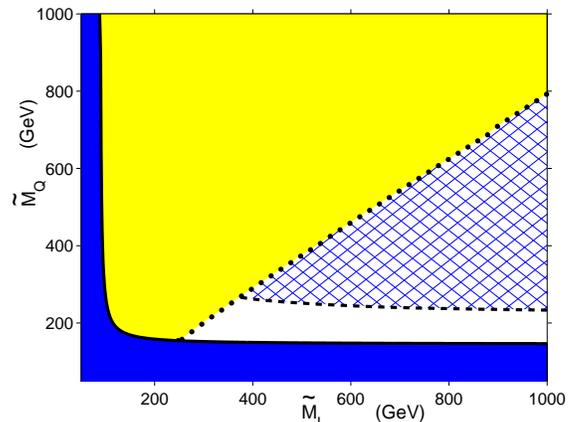}
\caption{\label{Fig2}Present and prospective constraints on masses for
left-handed sleptons
(${\tilde M}_L$) and  left-handed squarks (${\tilde M}_Q$). Dark shaded and
light shaded
regions are excluded by collider searches and charged current data,
respectively. Hatched
area would be excluded by a prospective
$+2.5\sigma$ deviation of $Q_W^{p}$ (10\%) or $Q_W^e$ (20\%) from the SM
prediction.}
\end{figure}

\end{document}